\documentclass[letterpaper,showpacs,superscriptaddress,twocolumn,aps,prl]{revtex4-1}
\usepackage{graphicx}
\begin{document}

\title{Efficient fault-tolerant decoding of topological color codes}
\author{Ashley M.~Stephens}\email{astephens@nii.ac.jp}
\affiliation{National Institute of Informatics, 2-1-2 Hitotsubashi, Chiyoda-ku, Tokyo 101-8430, Japan}

\date{\today}
\begin{abstract}
Topological color codes defined by the 4.8.8 semiregular lattice feature geometrically local check operators and admit transversal implementation of the entire Clifford group, making them promising candidates for fault-tolerant quantum computation. Recently, several efficient algorithms for decoding the syndrome of color codes were proposed. Here, we modify one of these algorithms to account for errors affecting the syndrome, applying it to the family of triangular 4.8.8 color codes encoding one logical qubit. For a three-dimensional bit-flip channel, we report a threshold error rate of 0.0208(1), compared with 0.0305(4) previously reported for an integer-program-based decoding algorithm. When we account for circuit details, this threshold is reduced to 0.00143(1) per gate, compared with 0.00672(1) per gate for the surface code under an identical noise model.
\end{abstract}

\pacs{03.67.Lx, 03.67.Pp}
\maketitle
\section{Introduction}
Fault-tolerant error correction promises to enable the reliable storage and manipulation of quantum information, but only if errors affecting qubits are not strongly correlated and occur with a probability below some threshold error rate \cite{Aliferis06,Gottesman1}. In practice, error correction should tolerate realistically high error rates and be compatible with experimentally feasible technology, without requiring excessive overhead in order to achieve a sufficiently reliable universal set of logical gates. In principle, topological error correction with the surface code appears to satisfy these criteria \cite{Kitaev2003,Bravyi2,Freedman1,Dennis2002}, enabling schemes that tolerate error rates of approximately one percent per gate for qubits constrained to a two-dimensional array with local interactions \cite{Raussendorf3,Raussendorf2007,Fowler1,Stephens13a}. However, the overhead of these schemes remains daunting \cite{Martinis2012,Devitt}, motivating the search for other approaches---for example, see Refs.~\cite{Bravyi12,Paetznick13,Jochym13,Gottesman13}. Here, we are interested in error correction using topological color codes \cite{Bombin1,Bombin2}, which have many similarities with the surface code but in some cases also admit transversal implementation of logical gates that span the Clifford group. This property may help to reduce the overhead required to achieve universality via state distillation \cite{BK05}, making these codes promising candidates for fault-tolerant error correction \cite{Landahl1,Fowler11}.

One critical aspect of error correction is decoding, which is the task of inferring from the error syndrome a set of corrections that will return the system to the code space. In the first instance, a practical decoding algorithm should be computationally efficient. It should also tolerate uncertainty in the syndrome due to errors---in other words, it should be fault tolerant. Efficient fault-tolerant decoding of the surface code can be achieved by solving a particular graph matching problem \cite{Dennis2002}, among other techniques \cite{Duclos-Cianci1,Duclos-Cianci2,Wootton1,Bravyi1}. However, the equivalent approach for topological color codes introduces a hypergraph matching problem for which no efficient algorithm is known, although approximate solutions can still be used \cite{Wang10b}. Recently, several efficient decoding algorithms for color codes have been proposed. These involve iteratively decoding using message passing \cite{Sarvepalli1} or relating the error syndrome to syndromes across multiple copies of the surface code \cite{Bombin12,Delfosse1}, which are then decoded using existing techniques. However, the performance of these algorithms remains unknown in the important, realistic case where the qubits used to measure the error syndrome are themselves affected by errors.

Here, we investigate the efficient fault-tolerant decoding of color codes. We begin with the decoding algorithm due to Delfosse \cite{Delfosse1}, which relates the hypergraph matching problem associated with color codes to a set of three graph matching problems with efficient solutions. We modify this algorithm to account for syndrome errors, applying it to the family of codes defined by the 4.8.8 semiregular lattice embedded on a plane with a triangular boundary. For a three-dimensional bit-flip channel, we report a threshold of 0.0208(1), compared with 0.0305(4) previously reported for an integer-program-based decoding algorithm \cite{Landahl1}. This result indicates that decoding of topological color codes by graph matching is a feasible approach in the context of fault-tolerant error correction. When we account for correlated errors introduced by the syndrome measurement circuits, the threshold is reduced to 0.00143(1) per gate. This threshold is higher than others reported for the color code \cite{Landahl1,Wang10b}, but significantly lower than the equivalent threshold for the surface code under an identical noise model \cite{Stephens13}. 

\section{Triangular 4.8.8 \\topological color codes}
A topological color code is defined by a three-colorable cubic (trivalent) graph embedded on a surface. The faces of a graph are three-colorable if every face can be assigned one of three colors such that no two faces that share an edge are the same color. Data qubits reside at the nodes of the graph. Each face is associated with two stabilizer generators \cite{Gottesman97}, or check operators, given by
\begin{equation}
S_X=\bigotimes_{i\in n(f)}X_i,~~~S_Z= \bigotimes_{j\in n(f)}Z_j,
\end{equation}
where $f$ is a face in the embedding, $n(f)$ is the set of qubits incident on $f$, and $X$ and $Z$ are the single-qubit Pauli operators. The code space is the simultaneous +1 eigenspace of the stabilizer generators.

\begin{figure}
\begin{center}
\includegraphics[scale=0.35]{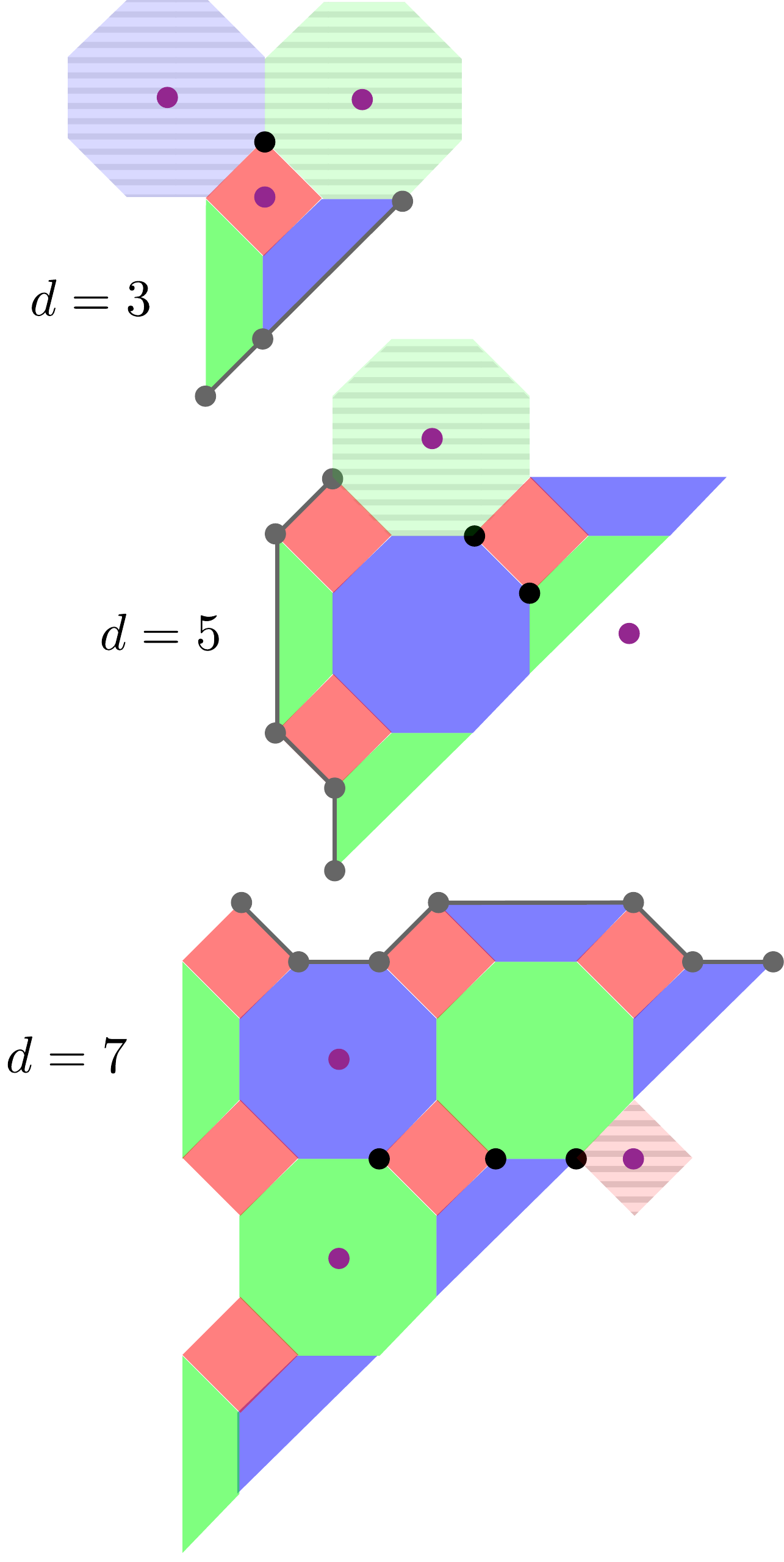}
\end{center}
\vspace{-10pt}
\caption{(Color online) Triangular 4.8.8 topological color codes of various distances $d$. Each node corresponds to a data qubit and each face corresponds to two stabilizer generators. Errors (black circles) anticommute with stabilizer generators (purple circles). Hatched faces indicate where errors terminate on the boundaries, but the implied qubits and stabilizer generators are not actually present. Logical Pauli operators (grey circles) are associated with the boundaries.}
\label{fig1}
\end{figure}

Here, we are particularly interested in the family of topological color codes defined by the 4.8.8 semiregular lattice embedded in a planar disc, which admit transversal implementation of the entire Clifford group \cite{Landahl1,Fowler11}. If we restrict the number of logical qubits to equal one, then we are left with a family of codes defined by lattices with triangular boundaries, as illustrated in Fig.~\ref{fig1}. The code parameters are $[[(d^2+2d-1)/2,1,d]]$ for odd-integer code distance $d$. Logical Pauli operators are connected chains of single-qubit Pauli operators along any one of the three boundaries. Additional logical qubits may be introduced by puncturing holes in the disc to modify its topology. In doing so, the stabilizer generators remain local in two spatial dimensions, ensuring compatibility with most quantum computer technologies \cite{Ladd2010}.

\section{Fault-tolerant decoding \\via graph matching}
In general, the error syndrome is obtained by measuring the eigenvalues of the complete set of stabilizer generators. However, for now, we will focus on the $X$-error syndrome, which is obtained by measuring the eigenvalues of only the $Z$-type stabilizer generators. The location of errors is implied by stabilizer generators whose eigenvalue is $-1$. For example, an isolated single-qubit error away from the boundaries will anticommute with the three adjacent stabilizer generators. However, more complicated chains of errors will only anticommute with the stabilizer generators at their terminals. Errors may also terminate on the boundaries, and so, in keeping with the three-coloration of the graph, each boundary is assigned a color complementary to the colors of the adjacent faces. Examples of errors and the corresponding syndromes are illustrated in Fig.~\ref{fig1}. Decoding is the procedure of inferring from the syndrome a set of corrections that will return the system to the code space. 

\begin{figure}
\begin{center}
\includegraphics[scale=0.35]{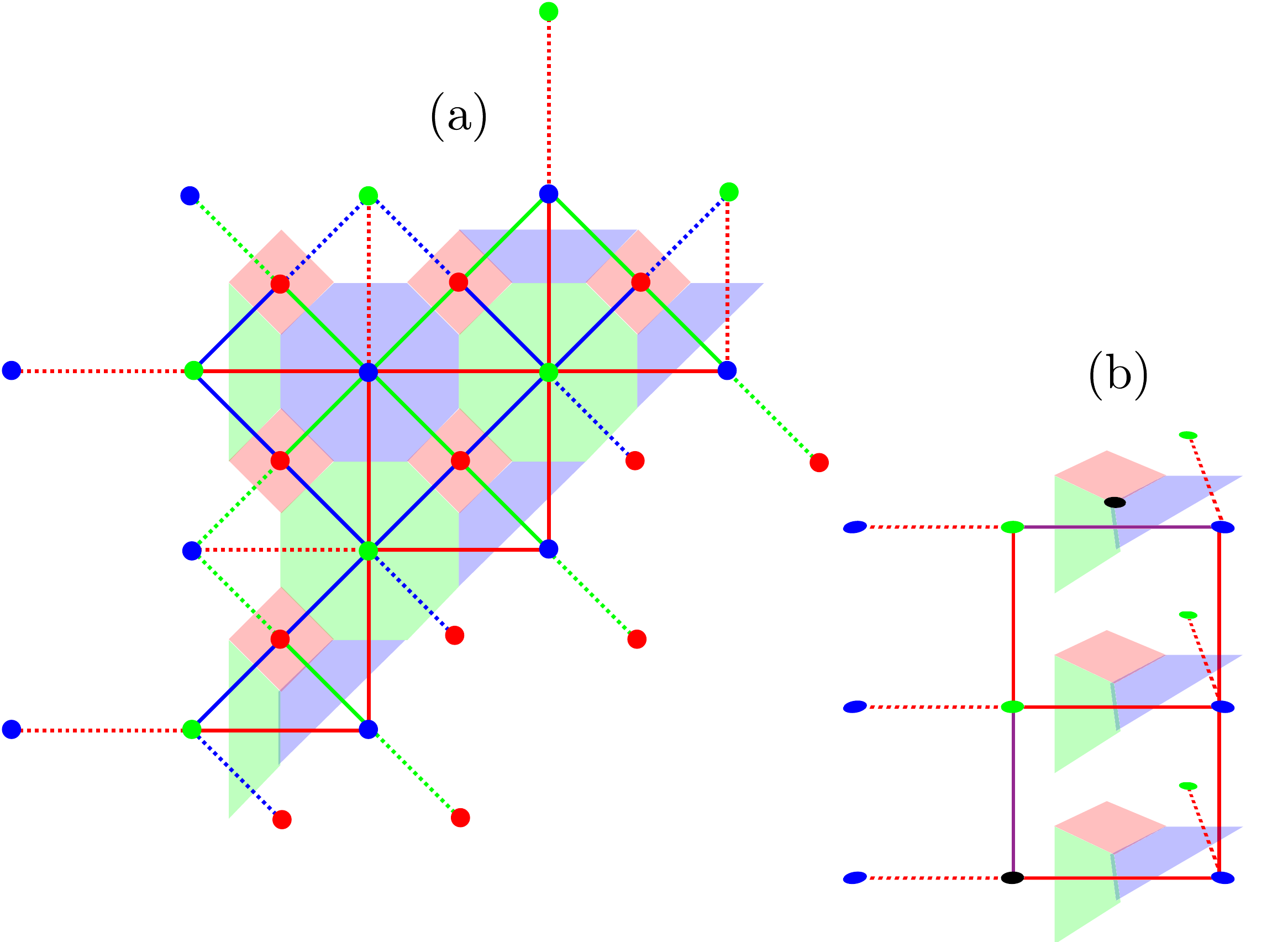}
\end{center}
\vspace{-10pt}
\caption{(Color online) (a) Dual graph of the triangular 4.8.8 topological color code for $d$=$7$. In contrast with Fig.~\ref{fig1}, each face corresponds to a data qubit and each node corresponds to a stabilizer generator. Nodes connected by dashed edges correspond to the boundaries. The dual graph is partitioned into three colored subgraphs that connect nodes of the complementary set of colors. (b) Red subgraph for $d$=$3$ where the syndrome is repeatedly measured in order from bottom to top. Errors affecting data qubits and syndrome qubits correspond to spacelike and timelike edges respectively.} 
\label{fig2}
\end{figure}

\begin{figure}
\begin{center}
\includegraphics[scale=0.35]{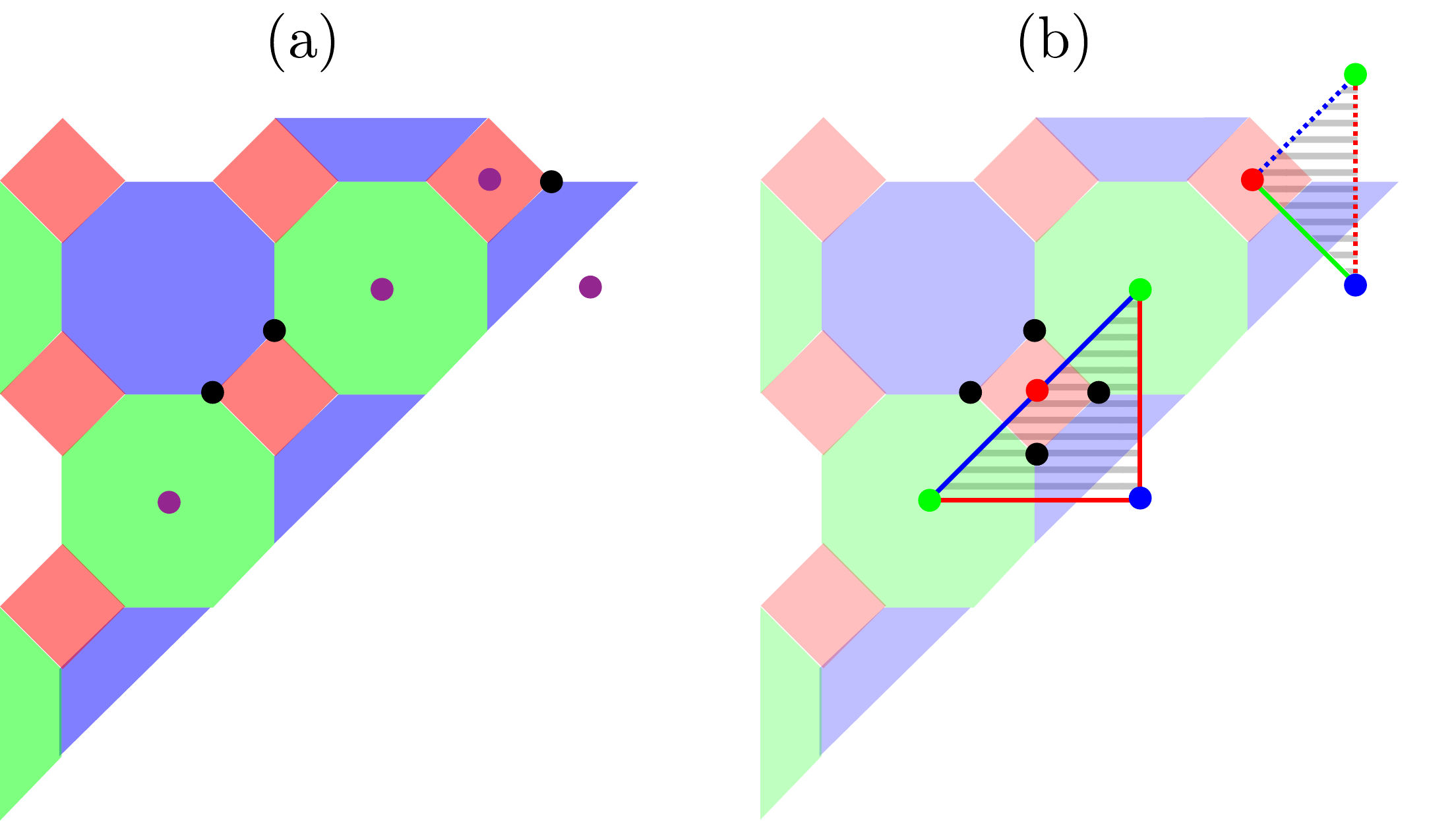}
\end{center}
\vspace{-10pt}
\caption{(Color online) {(a)} Set of errors and the corresponding syndrome. {(b)} Edges identified from each subgraph, the union of which partitions the data qubits into two sets. Corrections are applied to the three-qubit set (hatched region) to return the system to the code space. In this case, the combination of errors and corrections is equivalent to the identity operator.} 
\label{fig3}
\end{figure}

\begin{figure}
\begin{center}
\includegraphics[scale=0.35]{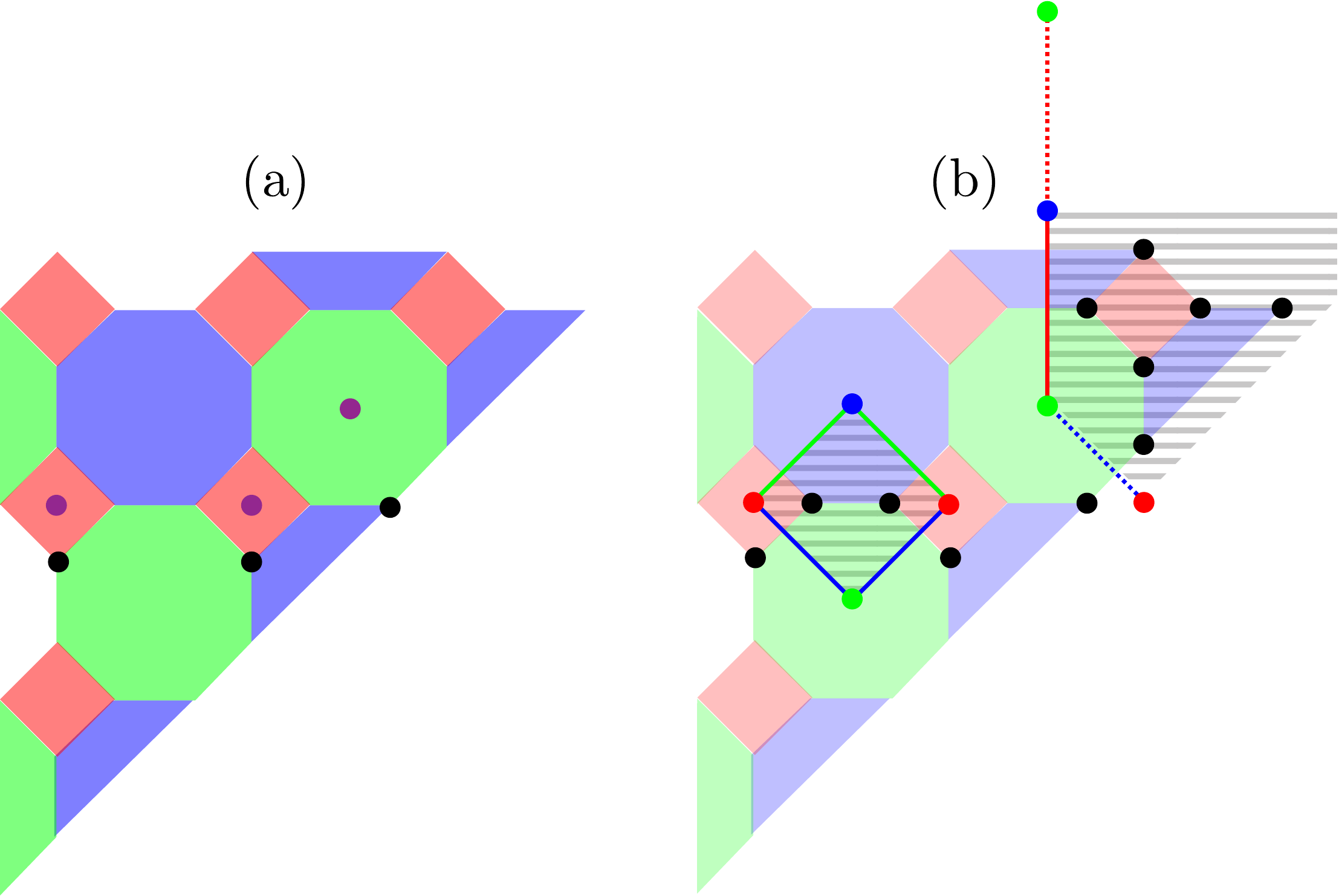}
\end{center}
\vspace{-10pt}
\caption{(Color online) {(a)} Set of errors and the corresponding syndrome. {(b)} Corrections are applied to the eight-qubit set (hatched region). In this case, the the combination of errors and corrections is equivalent to a logical operator.}
\label{fig4}
\end{figure}

Here, it is useful to introduce the dual graph, which can be partitioned into three colored subgraphs, as shown in Fig.~\ref{fig2}(a). Crucially, a single-qubit error will anticommute with only two adjacent stabilizer generators in each subgraph. In this way, when considered independently, each subgraph is equivalent to a surface code with its own syndrome \cite{Delfosse1}. The minimum-weight set of errors consistent with the syndrome of a surface code can be found efficiently by solving a particular graph matching problem using Edmonds' perfect matching algorithm \cite{Dennis2002,Wang10,Edmonds1,Kolmogorov1}. With this in mind, assuming that the syndrome is perfectly reliable, decoding proceeds as follows:
\begin{enumerate}
\item For each subgraph, find the minimum-weight set of errors consistent with the syndrome, corresponding to a set of edges.
\item Calculate the union of the edges identified in the three subgraphs, which partitions the data qubits into two sets.
\item Correct all qubits in the lowest-weight set.
\end{enumerate}
Figures \ref{fig3} and \ref{fig4} illustrate this procedure, giving examples of success and failure respectively. It may be possible to exploit correlations between the three subgraphs to more accurately identify errors \cite{Delfosse2}, but we have not investigated this possibility in detail.

In general, the error syndrome will be unreliable, as the circuits used to measure the eigenvalues of the stabilizer generators may introduce additional errors. This ambiguity is resolved by repeating the measurement circuits $d$ times. Then, the location of errors is implied by stabilizer generators whose eigenvalue is found to change between successive measurements. In particular, an error affecting a data qubit will cause the eigenvalues of the adjacent stabilizer generators to change from their previous values, while an error affecting a syndrome qubit will cause the eigenvalue of a single stabilizer generator to change (apparently) from its previous value and then change back again \cite{Dennis2002,Wang10}.

Again, our decoding strategy is to consider each subgraph as an independent surface code. However, to reflect the fact that the syndrome is repeatedly measured, each subgraph is extended along an axis which we identify with time \cite{Dennis2002,Wang10}. The minimum-weight set of errors consistent with the syndrome of each subgraph may include errors affecting data qubits and errors affecting syndrome qubits, which, respectively, correspond to spacelike and timelike edges in the subgraph, as illustrated in Fig.~\ref{fig2}(b). Therefore, we make the following modification to the second step of the decoding procedure:
\begin{itemize}
\item[2$^\prime$.] Calculate the two-dimensional projection of the spacelike edges identified in the three subgraphs, which partitions the data qubits into two sets. Edges cancel if they occur an even number of times in the projection.
\end{itemize}

\section{Syndrome measurement circuits}
Next, we specify circuits to measure the eigenvalues of the stabilizer generators. At the simplest level, standard single-qubit operator measurements are sufficient for this purpose, as the structure of the code limits the propagation of errors \cite{Landahl1}. However, more intricate circuits may be useful to trade off various parameters such as threshold, circuit depth, and connectivity \cite{Fowler11}. Here, circuits are chosen to ensure that high-weight correlated errors are relatively unlikely without introducing a stochastic verification procedure. 

\begin{figure}
\begin{center}
\includegraphics[scale=1.00]{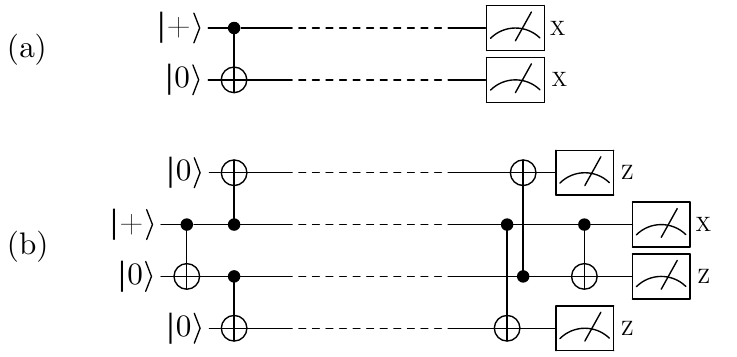}
\end{center}
\vspace{-12pt}
\caption{(Color online) (a) Circuit used to measure the eigenvalues of a weight-four stabilizer generator. The dashed wires indicate where each ancillary qubit interacts with two distinct data qubits. (b) Circuit used to measure the eigenvalues of a weight-eight stabilizer generator. The $Z$ measurements reveal any weight-two errors in the cat state caused by a single error occurring during its preparation \cite{DiVincenzo1}. To execute these circuits, assuming that qubits are constrained to a two-dimensional array, each qubit must be allowed to interact with either three or five of its neighbors.}
\label{fig5}
\end{figure}

Weight-four stabilizer generators are measured with a two-qubit cat state \cite{Shor1}, as shown in Fig.~\ref{fig5}(a). Weight-eight stabilizer generators are measured with a four-qubit cat state, which is prepared according to the left-hand side of Fig.~\ref{fig5}(b). In preparing this state, a single error may propagate to several errors that may in turn propagate to several data qubits. However, the circuit is designed so that it is possible to determine the nature of such errors with a simple post-processing operation \cite{DiVincenzo1}. Referring to the right-hand side of Fig.~\ref{fig5}(b), the $Z$ measurements detect these errors while the $X$ measurement reveals the eigenvalue of the stabilizer generator, as required. When correlated errors are detected, appropriate corrections are applied to the affected data qubits.

\section{Numerical results}
We simulate error correction with the color code using standard Monte Carlo methods \cite{Saito1}. In each instance we generate a set of errors based on some noise model, decode the corresponding syndrome, and apply corrections as required. We sample the logical error rate as a function of the physical error rate for odd code distances between 9 and 21. Following Refs.~\cite{Wang03,Landahl1}, a universal scaling ansatz accounting for finite-size effects is fit to the data and the threshold is extracted (where $R^2>0.999$ in all cases). Where appropriate, we quote the lowest of the separate thresholds for $X$ and $Z$ logical errors, which sets the overall threshold.

\begin{figure}
\begin{center}
\includegraphics[scale=0.5]{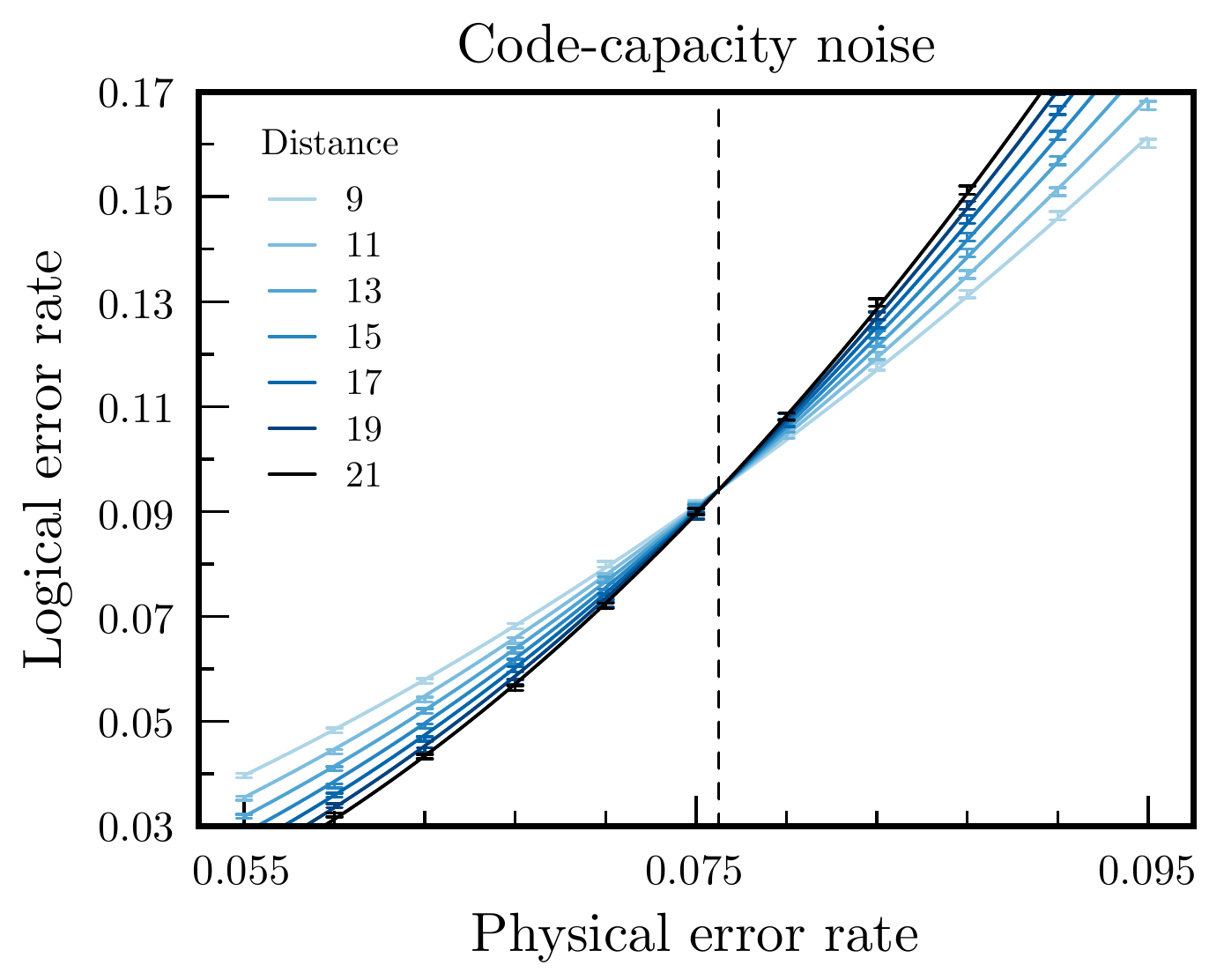}
\includegraphics[scale=0.5]{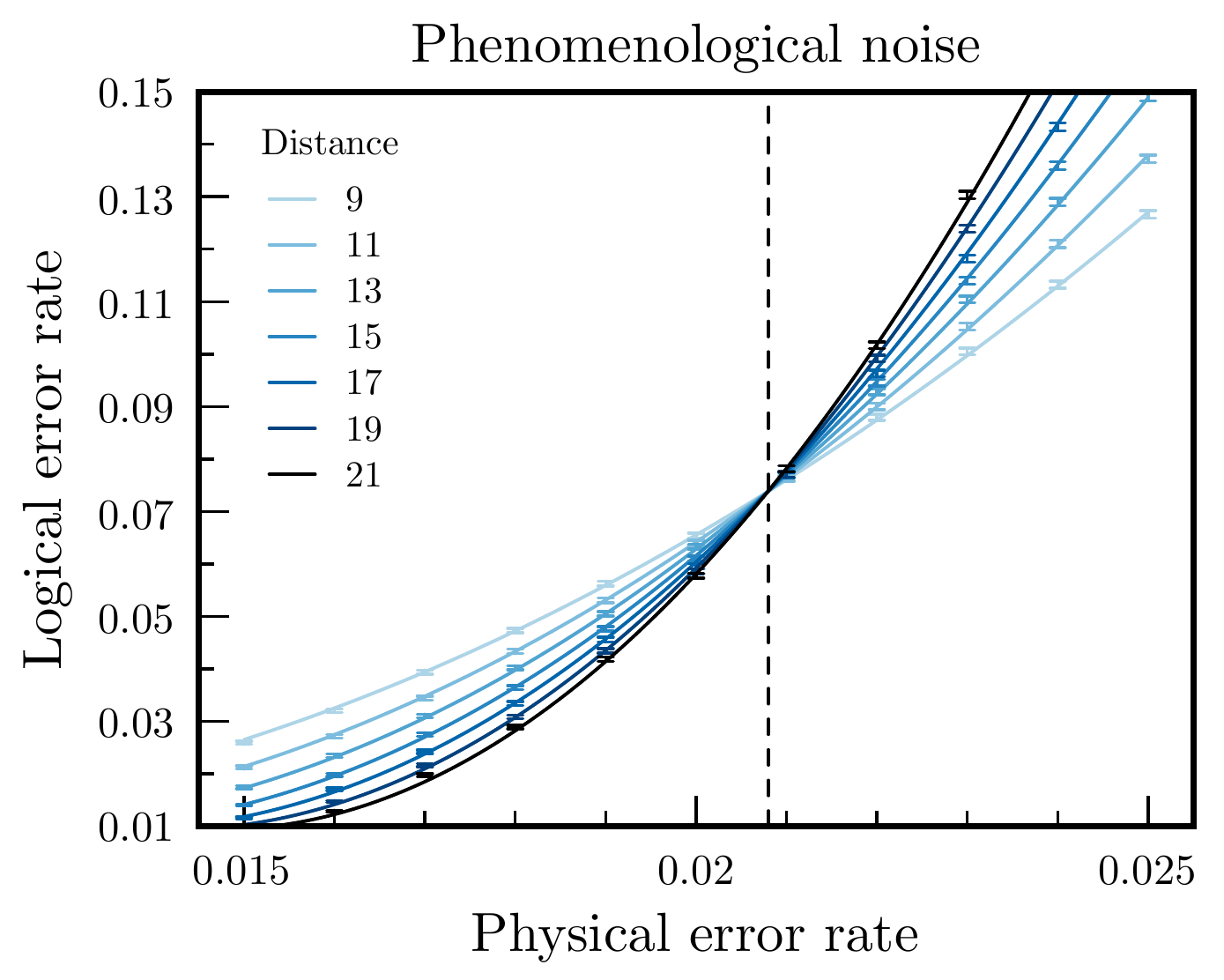}
\includegraphics[scale=0.5]{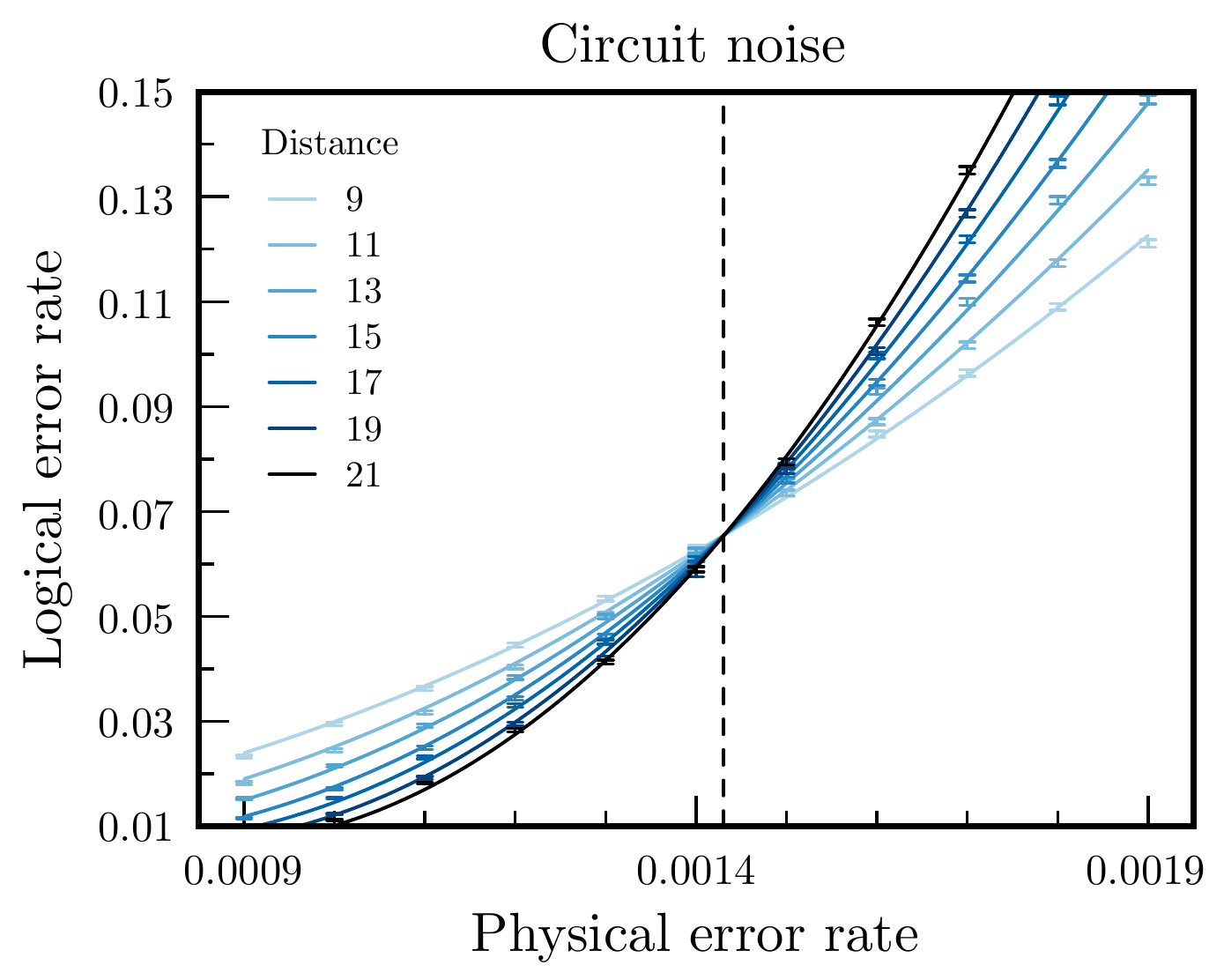}
\end{center}
\vspace{-16pt}
\caption{(Color online) Logical error rate as a function of physical error rate for various values of the code distance. Error bars indicate a $\pm2\sigma$ statistical error and curves are best fits accounting for finite-size effects \cite{Wang03,Landahl1}. The value of the physical error rate at the intersection (extracted from the fitting and indicated by the vertical dashed line) is the threshold error rate. Exact values are given in Table \ref{tab1}.}
\label{fig6}
\end{figure}

\begin{table}
\begin{center}
\vspace*{4pt}   
\begin{tabular}{llllc}
\multicolumn{5}{c}{Code-capacity noise}\\
\hline\hline
Lattice & Surface & Threshold & Decoding algorithm & Ref.\\
\hline
{\it 4.8.8} & {\it Triangle} & {\it 0.0760(2)} & {\it Graph matching$^*$} & {\it ---} \\
\hline
4.8.8 & Triangle & 0.1056(1) & Integer program & \cite{Landahl1} \\
4.8.8 & Torus & 0.087 & Renormalization$^*$ & \cite{Bombin12} \\
4.8.8 & Triangle & 0.0887 & Hypergraph matching & \cite{Wang10} \\
4.8.8 & Torus & 0.109(2) & Optimal & \cite{Katzgraber09} \\
4.8.8 & Torus & 0.10925(5) & Optimal & \cite{Ohzeki09} \\
\hline
6.6.6 & Torus & 0.078 & Renormalization & \cite{Sarvepalli1} \\
6.6.6 & Torus & 0.087 & Graph matching$^*$ & \cite{Delfosse1} \\
6.6.6 & Torus & 0.109(2) & Optimal & \cite{Katzgraber09} \\
6.6.6 & Torus & 0.1097(1) & Optimal & \cite{Ohzeki09} \\
\hline\hline
\end{tabular}
\begin{tabular}{llllc}
\\
\multicolumn{5}{c}{Phenomenological noise}\\
\hline\hline
Lattice & Surface & Threshold & Decoding algorithm & Ref.\\
\hline
{\it 4.8.8} & {\it Triangle} & {\it 0.0208(1)} & {\it Graph matching$^*$} & {\it ---} \\
\hline
4.8.8 & Triangle & 0.0305(4) & Integer program & \cite{Landahl1} \\
\hline
6.6.6 & Torus & 0.045(2) & Optimal & \cite{Andrist11} \\
\hline\hline
\end{tabular}
\begin{tabular}{llllc}
\\
\multicolumn{5}{c}{Circuit noise}\\
\hline\hline
Lattice & Surface & Threshold & Decoding algorithm & Ref.\\
\hline
{\it 4.8.8} & {\it Triangle} & {\it 0.00143(1)} & {\it Graph matching$^*$} & {\it ---} \\
\hline
4.8.8 & Triangle & 0.00082(3) & Integer program & \cite{Landahl1} \\
4.8.8 & Triangle & 0.001 & Hypergraph matching & \cite{Wang10} \\ 
\hline\hline
\end{tabular}
\vspace{2pt} 
\caption{Threshold error rates for various topological color codes defined by lattices embedded on surfaces, using various decoding algorithms, from references as indicated. An asterisk indicates that the decoding algorithm involves relating the error syndrome of the color code to syndromes across multiple copies of the surface code. Rows in italics are the present results, and no uncertainty is given if none was reported in the associated reference.}
\label{tab1}
\end{center}
\end{table}

Firstly, we consider {\it code-capacity noise}, where an $X$ error is applied to each data qubit with probability $p$. This noise model assumes that the syndrome is perfectly reliable. In this case, we observe a threshold at $p=0.0760(2)$.

Next, we consider {\it phenomenological noise}, where an $X$ error is applied to each data qubit and each syndrome bit value with probability $p$. This noise model captures the behavior of errors arising in the syndrome measurement circuits without explicitly considering their correlations. In this case, we observe a threshold at $p=0.0208(1)$.

Lastly, we consider {\it circuit noise}, where errors are applied to the individual gates in the syndrome measurement circuits. Specifically, each gate fails independently with probability $p$, where faulty preparation gives the incorrect eigenstate, a faulty gate introduces one- or two-qubit depolarizing noise, and faulty measurement reports the incorrect eigenstate. In this case, we observe a threshold at $p=0.00143(1)$. The equivalent threshold for the surface code under this noise model is $p=0.00672(1)$  \cite{Stephens13}. If we lower the probability of failure of identity gates to $p/10$ to reflect the fact that quantum memories may be significantly more reliable than other gates, then the threshold is increased to $p=0.00213(1)$.

Numerical data for the three noise models are shown in Fig.~\ref{fig6} and our results are compared with previously reported thresholds for a range of topological color codes and decoding algorithms in Table I.

\section{Summary and further work}
In summary, we have reported threshold error rates for the family of triangular 4.8.8 topological color codes, making use of an efficient fault-tolerant decoding algorithm based on graph matching. In the important case where one accounts for syndrome errors, our results accord well those in Refs.~\cite{Landahl1,Wang10b}, giving further evidence that thresholds for color codes are significantly lower than for the surface code. This would largely undermine the appealing features of color codes, namely their relatively high error-correcting rate and transversal implementation of the entire Clifford group. On the other hand, thresholds for color codes are still significantly higher than the highest thresholds reported for concatenated codes assuming that qubits constrained to a one- or two-dimensional array with local interactions \cite{Stephens09,Spedalieri09}. 

With respect to the decoding algorithm, there is still the prospect of improvement. For example, our simulations indicate that the full algebraic distance of the codes is not achieved. This is due to two reasons: Firstly, the syndrome measurement circuits introduce correlated errors that are not accounted for in the decoding algorithm. Secondly, the decoding algorithm fails for certain problematic error configurations, irrespective of these correlated errors. For example, Fig.~\ref{fig4} illustrates that fewer than $(d+1)/2$ errors may result in a logical error in some instances. Addressing these failings will be necessary to determine the optimal performance of topological color codes for fault-tolerant error correction. Nevertheless, our results indicate that decoding by graph matching is a feasible approach in this context. 

{\it Acknowledgements:} This work was supported by the FIRST Program in Japan. Thanks to W.~Munro and K.~Nemoto for offering helpful advice and to M. Carrasco for carefully reading the manuscript.

\bibliographystyle{unsrtnat}

\begin{thebibliography}{10}

\bibitem{Aliferis06}
P.~Aliferis, D.~Gottesman, and J.~Preskill, {\it Quantum Inf. Comput.} \textbf{6}, 97 (2006).

\bibitem{Gottesman1}
D.~Gottesman, in {\it Proc. Symp. Appl. Math.} {\bf 68}, 13 (Amer. Math. Soc., Providence, Rhode Island, 2010).

\bibitem{Kitaev2003}
A.~Y.~Kitaev, {\it Ann. Phys.} {\bf 303}, 2 (2003).

\bibitem{Bravyi2}
S.~B.~Bravyi and A.~Y.~Kitaev, arXiv:quant-ph/9811052.

\bibitem{Freedman1}
M.~H.~Freedman and D.~A.~Meyer, {\it Found. Comp. Math.} {\bf 1}, 325 (2001).

\bibitem{Dennis2002}
E.~Dennis, A.~Kitaev, A.~Landahl, and J.~Preskill, {\it J. Math. Phys.} {\bf 43}, 4452 (2002).

\bibitem{Raussendorf3}
R.~Raussendorf, J.~Harrington, and K.~Goyal, {\it New J. Phys.} {\bf 9}, 199 (2007).

\bibitem{Raussendorf2007}
R.~Raussendorf and J.~Harrington, {\it Phys. Rev. Lett.} {\bf 98}, 190504 (2007).

\bibitem{Fowler1}
A.~G.~Fowler, A.~M.~Stephens, and P.~Groszkowski, {\it Phys. Rev. A} \textbf{80}, 052312 (2009).

\bibitem{Stephens13a}
A.~M.~Stephens, W.~J.~Munro, and K.~Nemoto, {\it Phys. Rev. A} \textbf{88}, 060301(R) (2013).

\bibitem{Martinis2012}
A.~G.~Fowler, M.~Mariantoni, J.~M.~Martinis, and A.~N.~Cleland, {\it Phys. Rev. A} \textbf{86}, 032324 (2012).

\bibitem{Devitt}  
S.~J.~Devitt, A.~M.~Stephens, W.~J.~Munro, and K.~Nemoto, {\it Nat. Commun.} \textbf{4}, 2524 (2013).

\bibitem{Bravyi12}  
S.~Bravyi and J.~Haah, {\it Phys. Rev. A} {\bf 86}, 052329 (2012).

\bibitem{Paetznick13}
A.~Paetznick and B.~W.~Reichardt, {\it Phys. Rev. Lett.} {\bf 111}, 090505 (2013).

\bibitem{Jochym13} 
T.~Jochym-O'Connor and R.~Laflamme, {\it Phys. Rev. Lett.} {\bf 112}, 010505 (2014).

\bibitem{Gottesman13}
D.~Gottesman, arXiv:1310.2984.

\bibitem{Bombin1}
H.~Bombin and M.~A.~Martin-Delgado, {\it Phys. Rev. Lett.} {\bf 97}, 180501 (2006).

\bibitem{Bombin2}
H.~Bombin and M.~A.~Martin-Delgado, {\it Phys. Rev. A} {\bf 76}, 012305 (2007).

\bibitem{BK05}
S.~Bravyi and A.~Kitaev, {\it Phys. Rev. A} {\bf 71}, 022316 (2005).

\bibitem{Landahl1} 
A.~J.~Landahl, J.~T.~Anderson, and P.~R.~Rice, arXiv:1108.5738.

\bibitem{Fowler11}
A.~G.~Fowler, {\it Phys. Rev. A} {\bf 83}, 042310 (2011).

\bibitem{Duclos-Cianci1}  	
G.~Duclos-Cianci and D.~Poulin, {\it Phys. Rev. Lett.} {\bf 104}, 050504 (2010).

\bibitem{Duclos-Cianci2}  		
G.~Duclos-Cianci and D.~Poulin, {\it Quantum Inf. Comput.} {\bf 14}, 721 (2014).

\bibitem{Wootton1} 
J.~R.~Wootton and D.~Loss, {\it Phys. Rev. Lett.} {\bf 109}, 160503 (2012).

\bibitem{Bravyi1}  
S.~Bravyi and J.~Haah, {\it Phys. Rev. Lett.} {\bf 111}, 200501 (2013).

\bibitem{Wang10b}
D.~S.~Wang, A.~G.~Fowler, C.~D.~Hill, and L.~C.~L.~Hollenberg, {\it Quantum Inf. Comp.} {\bf 10}, 780 (2010).

\bibitem{Sarvepalli1}
P.~Sarvepalli and R.~Raussendorf, {\it Phys. Rev. A} {\bf 85}, 022317 (2012).

\bibitem{Bombin12}
H.~Bombin, G.~Duclos-Cianci, and D.~Poulin, {\it New J. Phys.} {\bf 14} (2012).

\bibitem{Delfosse1}
N.~Delfosse, {\it Phys. Rev. A} {\bf 89}, 012317 (2014).

\bibitem{Stephens13}
A.~M.~Stephens, arXiv:1311.5003.

\bibitem{Gottesman97}
D.~Gottesman, Ph.D. thesis, California Institute of Technology (1997).

\bibitem{Ladd2010}
T.~D.~Ladd, F.~Jelezko, R.~Laflamme, Y.~Nakamura, C.~Monroe, and J.~L.~O'Brien, {\it Nature} {\bf 464}, 45 (2010). 

\bibitem{Wang10}
D.~S.~Wang, A.~G.~Fowler, A.~M.~Stephens, and L.~C.~L.~Hollenberg, {\it Quantum Inf. Comput.} \textbf{10}, 456 (2010).

\bibitem{Edmonds1}
J.~Edmonds, {\it Canad. J. Math.}, {\bf 17}, 449 (1965).

\bibitem{Kolmogorov1}
V.~Kolmogorov, {\it Math. Program. Comput.} {\bf 1}, 43 (2009).

\bibitem{Delfosse2}
N.~Delfosse and J.-P.~Tillich, arXiv:1401.6975.

\bibitem{Shor1}
P.~W.~Shor, in {\it Proc. Symp. Found. Comp. Sci.} {\bf 37}, 56 (IEEE Comp. Soc., Los Alamitos, California, 1996).

\bibitem{DiVincenzo1}
D.~P.~DiVincenzo and P.~Aliferis, {\it Phys. Rev. Lett.} {\bf 98}, 020501 (2007).

\bibitem{Saito1}
M.~Saito and M.~Matsumoto, in {\it Monte Carlo and Quasi-Monte Carlo Methods 2006} (Springer, Berlin, 2008), Vol. 2, p. 607.

\bibitem{Wang03}
C.~Wang, J.~Harrington, and J.~Preskill, {\it Ann. Phys.} {\bf 303}, 31  (2003).

\bibitem{Katzgraber09}
H.~G.~Katzgraber, H.~Bombin, and M.~A.~Martin-Delgado, {\it Phys. Rev. Lett.} {\bf 103}, 090501 (2009).

\bibitem{Ohzeki09}
M.~Ohzeki, {\it Phys. Rev. E} {\bf 80}, 011141 (2009).

\bibitem{Andrist11}
R.~S.~Andrist, H.~G.~Katzgraber, H.~Bombin, and M.~A.~Martin-Delgado, {\it New J. Phys.} {\bf 13}, 083006 (2011).

\bibitem{Stephens09}
A.~M.~Stephens and Z.~W.~E.~Evans, {\it Phys. Rev. A} \textbf{80}, 022313 (2009).

\bibitem{Spedalieri09}
F.~M.~Spedalieri and V.~P.~Roychowdhury, {\it Quantum Inf. Comput.} \textbf{9}, 666 (2009).

\end{thebibliography}

\end{document}